\documentclass[pre,twocolumn]{revtex4-1}
\usepackage{amsmath}
\usepackage{graphicx}
\usepackage{color}

\begin{document}

\title{On elastic constants of zero-temperature amorphous solids}
\author{Grzegorz Szamel}
\affiliation{Department of Chemistry,
Colorado State University, Fort Collins, CO 80523}

\begin{abstract}
  Elastic constants of zero-temperature amorphous solids are given
  as the difference between the Born term, which results from a hypothetical affine
  deformation of an amorphous solid, and a correction term which originates
  from the fact that the deformation of an amorphous solid due to an applied
  stress is, at the microscopic level,
  non-affine. Both terms are non-negative and thus it is \textit{a priori} not obvious
  that the resulting elastic constants are non-negative.
  In particular, theories that approximate the correction term may
  spuriously predict negative elastic constants and thus an instability
  of an amorphous solid. 
  Here we derive alternative expressions for elastic constants of
  zero-temperature amorphous solids that are explicitly non-negative.
  These expressions provide a useful blueprint for approximate theories
  for elastic constants and sound damping in zero temperature amorphous solids.
\end{abstract}

\maketitle

\section{Introduction}\label{intro}

Elastic constants of zero-temperature crystalline solids consisting of
particles with pair-wise additive interactions and with one atom per unit cell
are given by the well-known Born formulas \cite{BornHuang}. However, already
for crystalline solids with more than one atom per unit cell the elastic constants
are given as the difference between the Born term and a correction term originating
from an internal relaxation within the unit cell upon deformation \cite{Lutsko1989}.
The correction term becomes more important for amorphous solids which are completely
devoid of a crystalline lattice \cite{LeonfortePRB,LemaitreJSP}. The correction term
is rationalized in terms of the so-called \textit{non-affine displacement field},
which originates from the inherent disorder of amorphous solids
\cite{LeonfortePRB,LemaitreJSP,Wittmer2002,Tanguy2002}. 
In the following we will refer to the two terms in the standard expression
for the elastic constants of zero-temperature amorphous solids
\cite{LemaitreJSP,KarmakarLP2010} as the Born term and
the correction or non-affine term.

Both the Born term and the non-affine term
are positive definite. Thus, it is \textit{a priori} not obvious whether their
difference is non-negative, as elastic constants should be. This problem becomes
important if an approximate theory is used to calculate the correction
term. If such a theory overestimates the magnitude of the correction term,
it may predict a spurious instability of the amorphous solid in question.

We recall that a similar situation occurred in the context of the description
of the dynamics of stochastic systems, \textit{e.g.} overdamped colloidal
suspensions and kinetically constrained models. In this case, interaction-induced
or constraint-induced
change of a relaxation rate was initially described in terms of a correction term that
was \textit{subtracted} from the relaxation rate of the non-interacting system. 
Approximate theories for the correction term often 
resulted in qualitatively incorrect description of the relaxation \cite{HessKlein}
or predicted spurious dynamic transitions \cite{JaeckleEisinger}.
This situation got resolved after the formal expression for the
relaxation rate was re-written in terms of a new quantity called the
irreducible memory function  \cite{CHess,Kawasaki}, resulting in an expression
that was explicitly non-negative. This new formulation
lead to useful approximate descriptions of strongly interacting colloidal
suspensions  \cite{SzamelLoewen}.

Here we achieve a similar goal for the elastic constants of zero-temperature
amorphous solids.
We derive new, alternative expressions for elastic constants that are
explicitly non-negative. The elastic constants are expressed as
a ratio of the Born term squared
and a sum of the Born term and a new non-affine correction term, which
is positive definite.
We expect that the new expressions will allow for a reformulation
of existing approximate descriptions of sound propagation in amorphous solids.

\section{Statement of the problem}\label{problem}

We consider an amorphous zero-temperature solid consisting of $N$ particles
interacting \textit{via} a spherically symmetric pair-wise additive potential.
We assume that particles' positions correspond to a local minimum of the
potential energy, \textit{i.e.} an inherent structure \cite{SastryDeBStil}.
We are interested in small displacements of the particles from their
inherent structure positions. The quantity that describes the response of the
system to external perturbations is the Hessian,
\begin{eqnarray}\label{H}
  \mathcal{H}_{il} = - \frac{\partial \mathbf{F}_i(\{ \mathbf{R}_m \})}
          {\partial \mathbf{R}_l},
\end{eqnarray}
where $\mathbf{R}_i$, $i=1,\ldots,N$ denote the inherent structure positions
of the particles and $\mathbf{F}_i(\{ \mathbf{R}_m \})$ is the total force acting
on particle $i$,
\begin{eqnarray}\label{F}
  \mathbf{F}_i(\{ \mathbf{R}_m \}) = -\frac{\partial}{\partial \mathbf{R}_i} \sum_{j\neq i}
  V(R_{ij}),
\end{eqnarray}
with $R_{ij}=|\mathbf{R}_{ij}|\equiv|\mathbf{R}_i-\mathbf{R}_j|$ being the
interparticle distance and $V(r)$ being the pair potential.

We note that each element $\mathcal{H}_{ij}$ is a 3x3 tensor; if needed, we will
refer to the components of $\mathcal{H}_{ij}$ and other tensors using Greek indices,
\textit{e.g.} $\mathcal{H}_{i\alpha j\beta}=
- \frac{\partial F_{i\alpha}(\{ \mathbf{R}_m \})}{\partial R_{j\beta}}$.

Translational invariance implies that uniform displacements from the
inherent structure positions do not induce a restoring force. In other words,
the Hessian matrix has three linearly independent eigenvectors corresponding
to zero eigenvalue. The components of these eigenvectors do not depend on the
particles' positions and the eigenvectors can be chosen to point along directions
of the coordinate system,
\begin{eqnarray}\label{zeroE}
  \boldsymbol{\mathcal{E}}^{\alpha}_{0i} = N^{-1/2} \hat{\boldsymbol{\alpha}},
\end{eqnarray}
where $\hat{\boldsymbol{\alpha}}$ is a unit vector along the $\alpha$ axis.

In principle, there might be additional eigenvectors corresponding to zero
eigenvalue, \textit{e.g.} they would appear if our amorphous solid consisted
of disconnected clusters.
We assume that the three eigenvectors $\boldsymbol{\mathcal{E}}^{0\alpha}$ are the only
eigenvectors of the Hessian corresponding to zero eigenvalue. In other words,
we assume that the amorphous solid that we consider is stable. 

Furthermore, we will assume that the solid is, on average, isotropic.
For this reason, it has only two independent elastic constants or, equivalently,
two independent speeds of sound. To simplify the notation we formulate
our approach as the theory for the speeds of sound. The actual
elastic constants, the bulk modulus and the shear modulus, can be
easily obtained from the speeds of sound. 

To derive expressions for the speeds of sound,
we consider the following question: if we apply an external periodic 
force on the particles, what will be the resulting
displacement field in the limit of long-wavelengths?

To be more specific, we assume that at initial time $t=0$ we turn on a periodic force
acting on the particles. Force on particle $i$ is given by 
\begin{eqnarray}\label{extforce}
  \mathbf{f}_i = \hat{\mathbf{e}} e^{-i\mathbf{k}\cdot\mathbf{R}_i},
\end{eqnarray}
where $\hat{\mathbf{e}}$ is a unit vector that is parallel,
$\hat{\mathbf{e}}_L=\hat{\mathbf{k}}$,
or perpendicular, $\hat{\mathbf{e}}_T$, to
wavevector $\mathbf{k}$ for parallel (bulk) or transverse (shear) perturbations.
Note that we assumed a unit amplitude of the force; since we are working
in the harmonic approximation, the strength of the force does not play any role.

As a result of external force \eqref{extforce} the particles get displaced
from their inherent structure positions. To mimic the procedure used in
computer simulations (sample deformation followed by relaxation \cite{OHern2003}),
we assume that after the force is turned on,
displacements $\mathbf{u}_i$ evolve according to 
overdamped dynamics with relaxation time $\tau$,
\begin{eqnarray}\label{Hexteom}
  \tau \partial_t \mathbf{u}_i(t) = -\sum_j \mathcal{H}_{ij} \cdot \mathbf{u}_j(t)
  + \mathbf{f}_i .
\end{eqnarray}
Here $\mathbf{u}_i$ is the displacement of particle $i$ from its inherent
structure position $\mathbf{R}_i$.

The dynamics described by Eqs. \eqref{Hexteom} is not the real dynamics
of the system. However, it is a useful auxiliary process that allows us to reach
the final state of the deformed system. 

We are interested in the long-time limit of the displacements. We will
analyze the evolution defined by Eqs. \eqref{Hexteom} in the Laplace space,
\begin{eqnarray}\label{Hexteomz}
  z \tau \mathbf{u}_i(z) = -\sum_j \mathcal{H}_{ij} \cdot \mathbf{u}_j(z)
  + \mathbf{f}_i/z.
\end{eqnarray}
Writing Eq. \eqref{Hexteomz} we used the fact that before the force was turned on,
the solid was un-deformed, \textit{i.e.} $\mathbf{u}_i(t=0)=0$.

The formal solution of Eq. \eqref{Hexteomz} reads
\begin{eqnarray}\label{Hextsol}
  \mathbf{u}_i(z) = \sum_j \left[z\tau +\mathcal{H}\right]^{-1}_{ij} \cdot \mathbf{f}_j/z.
\end{eqnarray}

In the small $z$ limit, \textit{i.e.} for $z\tau \ll 1$,
the displacement field is given by 
\begin{eqnarray}\label{Hextsolsmallz}
  \mathbf{u}_i(z\to 0) = \sum_j \left[\mathcal{H}\right]^{-1}_{ij} \cdot \mathbf{f}_j/z.
\end{eqnarray}
Thus, the long-time limits of the displacements $\mathbf{u}_i(t\to\infty)$ satisfy
the following equations,
\begin{eqnarray}\label{Hext}
  \mathbf{u}_i(t\to\infty) = \sum_j \left[\mathcal{H}\right]^{-1}_{ij} \cdot \mathbf{f}_j.
\end{eqnarray}
Eqs. \eqref{Hext} express force balance after deformation and they may have been
written directly. However, we find it convenient to use Eqs. \eqref{Hexteomz}
as the starting point of our analysis. 

As discussed above, the real microscopic displacements are in general non-affine.
However, on the basis of the macroscopic theory of elasticity, we expect
that after force \eqref{extforce} was applied, in the limit of small magnitude
of the wavevector $k=|\mathbf{k}|$ there will
be an affine component of the displacements that will be linearly related to the
force, with the coefficient of proportionality that is proportional to the inverse of the
product of the square of the speed of sound and the square of the wavevector $\mathbf{k}$.
In the small z limit the affine component of the displacement field will be given by
\begin{eqnarray}\label{affdispl0}
  \mathbf{u}_i^\text{aff}(z) = \left( k^2 c^2 \right)^{-1}\mathbf{f}_i/z.
\end{eqnarray}
To compare with relations derived from microscopic considerations we re-write
Eq. \eqref{affdispl0} as follows
\begin{eqnarray}\label{affdispl}
  k^2 c^2 \mathbf{u}_i^\text{aff}(z) = \mathbf{f}_i/z.
\end{eqnarray}
Depending on whether the external force is
parallel of perpendicular to the wavevector $\mathbf{k}$, one should use in
Eqs. (\ref{affdispl0}-\ref{affdispl})
the longitudinal, $c_L$, or the transverse, $c_T$, speed of sound. 

Our goal is to derive, in the limit of $z\tau\to 0$, 
a microscopic version of Eq. \eqref{affdispl}. In this way we will obtain
expressions for the speeds of sound. In the next section
we re-derive the standard expression and in the following section we derive
an alternative expression that is explicitly positive definite. 

\section{Re-derivation of the standard
expression for speeds of sound}\label{standard}

We define projection operator $\mathcal{P}$ that selects the affine part of
the displacement field and orthogonal projection $\mathcal{Q}$,
\begin{eqnarray}\label{Pdef}
  \mathbf{u}_i^\text{aff} = \mathcal{P}\mathbf{u}_i = 
  e^{-i\mathbf{k}\cdot\mathbf{R}_i}\hat{\mathbf{e}}
  \frac{1}{N}\sum_j e^{i\mathbf{k}\cdot\mathbf{R}_j}  \hat{\mathbf{e}}\cdot\mathbf{u}_j,
\end{eqnarray}
\begin{eqnarray}\label{Qdef}
  \mathcal{Q}\mathbf{u}_i = \mathbf{u}_i- \mathcal{P}\mathbf{u}_i.
\end{eqnarray}

Next, we apply $\mathcal{P}$
to both sides of Eq. \eqref{Hexteomz} and we also insert $\mathcal{P} + \mathcal{Q} = 1$
between the Hessian and the displacement field,
\begin{eqnarray}\label{HextP1}
  z\tau\mathcal{P}\mathbf{u}_i(z) &=& -\sum_j \mathcal{P} \mathcal{H}_{ij} \cdot
  \left(\mathcal{P}+\mathcal{Q}\right)\mathbf{u}_j(z)
  + \mathcal{P}\mathbf{f}_i/z
  \nonumber \\ &\equiv & -\sum_j \mathcal{P} \mathcal{H}_{ij} \cdot
  \left(\mathcal{P}+\mathcal{Q}\right)\mathbf{u}_j(z)
  + \mathbf{f}_i/z.
  \nonumber \\ 
\end{eqnarray}
Then, we apply apply $\mathcal{Q}$
to both sides of Eq. \eqref{Hexteomz} and again we insert the sum of $\mathcal{P}$ and
$\mathcal{Q}$ between the Hessian and the displacement field,
\begin{eqnarray}\label{HextQ1}
  z\tau\mathcal{Q}\mathbf{u}_i(z) = -\sum_j \mathcal{Q} \mathcal{H}_{ij} \cdot
  \left(\mathcal{P}+\mathcal{Q}\right)\mathbf{u}_j(z) .
\end{eqnarray}

Finally, we formally solve Eq. \eqref{HextQ1} for $\mathcal{Q} \mathbf{u}_i(z)$, 
substitute the result into Eq. \eqref{HextP1} and in this way we obtain
\begin{eqnarray}\label{avedispl1}
  && \left\{z\tau + \mathcal{P} \mathcal{H}\mathcal{P}
  - \mathcal{P} \mathcal{H}\mathcal{Q}
  \frac{1}{z\tau + \mathcal{Q} \mathcal{H} \mathcal{Q}}
  \mathcal{Q}\mathcal{H}\mathcal{P} \right\} \cdot \mathcal{P}\mathbf{u}
  = \mathbf{f}/z,
  \nonumber \\
\end{eqnarray}
where to simplify the notation we omitted summations over particles and
indices of the Hessian, displacement and force fields. 

To recover standard expressions for the speeds of sound we need to investigate
the small wavevector limit of the second and third term at the left-hand-side
of Eq. \eqref{avedispl1} and then take the small $z\tau$ limit. The
sum of the second and third terms will give the product of the squares of the wavevector
and the speed of sound, see Eq. \eqref{affdispl}.

Restoring summation over
particles and the indices, we can analyze the second term at the
left-hand-side of \eqref{avedispl1} as follows,
\begin{eqnarray}\label{PHP1}
  && \sum_j \mathcal{P} \mathcal{H}_{ij}\mathcal{P} \cdot \mathcal{P} \mathbf{u}_j
  \\ \nonumber &=& 
  \frac{1}{N^2}\hat{\mathbf{e}} e^{-i\mathbf{k}\cdot\mathbf{R}_i}\sum_{m,l}
  \hat{\mathbf{e}}\cdot\mathcal{H}_{lm}\cdot\hat{\mathbf{e}}
  e^{-i\mathbf{k}\cdot\mathbf{R}_{lm}}
  \sum_j e^{i\mathbf{k}\cdot\mathbf{R}_j} \hat{\mathbf{e}}\cdot\mathbf{u}_j
  \\ \nonumber &=&
  \left[\frac{k^2}{2N}
    \sum_l \sum_{m\neq l}
    \hat{\mathbf{e}}\cdot\frac{\partial^2 V(R_{lm})}{\partial \mathbf{R}_l^2}
    \cdot\hat{\mathbf{e}}\left(\hat{\mathbf{k}}\cdot\mathbf{R}_{lm}\right)^2
    + o(k^2) \right] \mathcal{P} \mathbf{u}_i,
\end{eqnarray}
where we used the fact that the zeroth-order in $k$ term vanishes due to
translational symmetry and the first-order term vanishes due to $i\leftrightarrow j$
symmetry.

For an isotropic system tensorial quantity
\begin{eqnarray}\label{PHP2}
  \mathcal{V}(\hat{\mathbf{k}}) = \frac{1}{2N}\sum_l \sum_{m\neq l}
    \frac{\partial^2 V(R_{lm})}{\partial \mathbf{R}_l^2}
    \left(\hat{\mathbf{k}}\cdot\mathbf{R}_{lm}\right)^2
\end{eqnarray}
consists of two independent components, longitudinal and transverse. These
components are proportional to Born approximations for the longitudinal and
transverse speed of sound squared,
\begin{eqnarray}\label{PHP3}
  \mathcal{V}(\hat{\mathbf{k}}) 
  = c_{LB}^2 \hat{\mathbf{k}}\hat{\mathbf{k}}
    +c_{TB}^2 \left(1-\hat{\mathbf{k}}\hat{\mathbf{k}}\right)
\end{eqnarray}
where $c_{LB}$ and $c_{TB}$ are given by 
\begin{eqnarray}\label{PHP4}
  c_{LB}^2 = \frac{1}{2N} \sum_i \sum_{j\neq i} 
  \hat{\mathbf{k}}\cdot \frac{\partial^2 V(R_{ij})}{\partial \mathbf{R}_i^2}
    \cdot \hat{\mathbf{k}}
    \left(\hat{\mathbf{k}}\cdot\mathbf{R}_{ij}\right)^2
\end{eqnarray}
and
\begin{eqnarray}\label{PHP5}
  && c_{TB}^2
  = \frac{1}{4N}
  \left[\boldsymbol{1} - \hat{\mathbf{k}}\hat{\mathbf{k}}\right] :
  \sum_i \sum_{j\neq i} 
  \frac{\partial^2 V(R_{ij})}{\partial \mathbf{R}_i^2}
  \left(\hat{\mathbf{k}}\cdot\mathbf{R}_{ij}\right)^2.
  \nonumber \\
\end{eqnarray}
Combining Eqs. (\ref{PHP1},\ref{PHP2},\ref{PHP3})
we get the following expression for the small wavevector limit
of the second term at the left-hand-side of Eq. \eqref{avedispl1},
\begin{eqnarray}\label{PHPfin}
&& \sum_j \mathcal{P} \mathcal{H}_{ij}\mathcal{P} \cdot \mathcal{P} \mathbf{u}_j
\\ \nonumber &=&
\left[c_{LB}^2 k^2 \left(\hat{\mathbf{k}}\cdot\hat{\mathbf{e}}\right)^2
    +c_{TB}^2 k^2 \left(1-\left(\hat{\mathbf{k}}\cdot\hat{\mathbf{e}}\right)^2\right)
    \right] \mathcal{P} \mathbf{u}_i,
\end{eqnarray}

The analysis of the third term at the left-hand-side of Eq. \eqref{avedispl1}
is a bit more tedious; it is presented in Appendix \ref{appA}.
The final result for the small wavevector and the subsequent small $z\tau$
limit of the third term is 
\begin{eqnarray}\label{QHQfin}
  && - \sum_{j,l,m} \mathcal{P} \mathcal{H}_{il}\mathcal{Q}
  \left[z\tau+ \mathcal{Q} \mathcal{H} \mathcal{Q} \right]_{lm}^{-1}
  \mathcal{Q}\mathcal{H}_{mj}\mathcal{P} \cdot \mathcal{P}\mathbf{u}_j
  \\ \nonumber &=&
  \left[\Delta c_{L}^2 k^2 \left(\hat{\mathbf{k}}\cdot\hat{\mathbf{e}}\right)^2
    +\Delta c_{T}^2 k^2 \left(1-\left(\hat{\mathbf{k}}\cdot\hat{\mathbf{e}}\right)^2\right)
    \right] \mathcal{P} \mathbf{u}_i,
\end{eqnarray}
where the contributions to the speeds of sound due to non-affine effects,
$\Delta c_{L}$ and $\Delta c_{T}$,
can be written in terms of tensorial field $\mathcal{W}_j$,
\begin{eqnarray}\label{Wa}
  \mathcal{W}_j(\hat{\mathbf{k}})
  = \sum_{l\neq j} \frac{\partial^2 V(R_{jl})}{\partial \mathbf{R}_j^2}
  \hat{\mathbf{k}}\cdot\mathbf{R}_{jl},
\end{eqnarray}
that quantifies the magnitude of the non-affine response, see Appendix \ref{appA}.
The expressions for $\Delta c_{L}^2$ and $\Delta c_{T}^2$ read
\begin{eqnarray}\label{QHQ3a}
  \Delta c_{L}^2 =
  - \hat{\mathbf{k}}\cdot \frac{1}{N} \sum_{l,m} 
  \mathcal{W}_l(\hat{\mathbf{k}})\cdot
  \left[ \mathcal{H}\right]^{-1}_{lm}\cdot
  \mathcal{W}_m(\hat{\mathbf{k}})\cdot \hat{\mathbf{k}},
  \nonumber \\
\end{eqnarray}
\begin{eqnarray}\label{QHQ4a}
  \Delta c_{T}^2 =
  - \frac{1}{2}
  \left[\boldsymbol{1} - \hat{\mathbf{k}}\hat{\mathbf{k}}\right] :
  \frac{1}{N} \sum_{l,m} 
  \mathcal{W}_l(\hat{\mathbf{k}})\cdot
  \left[ \mathcal{H}\right]^{-1}_{lm}\cdot
  \mathcal{W}_m(\hat{\mathbf{k}}).
  \nonumber \\
\end{eqnarray}
As discussed in Appendix \ref{appA}, terms
$\sum_{l,m}\mathcal{W}_l(\hat{\mathbf{k}})\cdot\left[ \mathcal{H}\right]^{-1}_{lm}\cdot
\mathcal{W}_m(\hat{\mathbf{k}})$ in Eqs. (\ref{QHQ3a}-\ref{QHQ4a})
should be understood as
$\sum_l \mathcal{W}_l(\hat{\mathbf{k}})\cdot\mathcal{U}_l(\hat{\mathbf{k}})$,
where $\mathcal{W}_l(\hat{\mathbf{k}})=\sum_m \mathcal{H}_{lm}\cdot
\mathcal{U}_m(\hat{\mathbf{k}})$. The latter equation has a unique solution
since $\mathcal{W}_m(\hat{\mathbf{k}})$ is orthogonal to all eigenvectors
of the Hessian corresponding to zero eigenvalue. 
We note that since the Hessian is non-negative definite,
contributions (\ref{QHQ3a}-\ref{QHQ4a}) are negative-definite.

Combining Eq. \eqref{avedispl1} and Eqs. (\ref{PHPfin}-\ref{QHQfin}) and taking
the $z\tau\to 0$ limit we recover relation \eqref{affdispl}
between the affine component of the displacement field and the external force,
with speeds of sound squared given by
\begin{eqnarray}\label{c2long1}
  && c_{L}^2 =
   c_{LB}^2  -\frac{1}{N} \sum_{l,m} 
  \hat{\mathbf{k}}\cdot\mathcal{W}_l(\hat{\mathbf{k}})\cdot
  \left[ \mathcal{H}\right]^{-1}_{lm}\cdot
  \mathcal{W}_m(\hat{\mathbf{k}})\cdot\hat{\mathbf{k}},
  \nonumber \\
\end{eqnarray}
\begin{eqnarray}\label{c2trans1}
  && c_{T}^2 =
  c_{TB}^2  -\frac{1}{N} \sum_{l,m} 
  \hat{\mathbf{e}}_T\cdot\mathcal{W}_l(\hat{\mathbf{k}})\cdot
  \left[ \mathcal{H}\right]^{-1}_{lm}\cdot
  \mathcal{W}_m(\hat{\mathbf{k}})\cdot\hat{\mathbf{e}}_T.
  \nonumber \\
\end{eqnarray}
Expressions (\ref{c2long1}-\ref{c2trans1}) for the speeds of sound squared
are equivalent to the
standard expressions for elastic constants of zero-temperature amorphous solids
\cite{Lutsko1989,LemaitreJSP,KarmakarLP2010}. As discussed earlier,
expressions (\ref{c2long1}-\ref{c2trans1}) are not explicitly non-negative
and approximate theories for the second terms in these expressions may lead
to spurious predictions of instabilities.

\section{Alternative expression for speeds of sound}\label{alter}

We now derive alternative expressions for speeds of sound squared which
makes their non-negative property explicit.

We start by re-writing Eq. \eqref{HextP1} as follows,
\begin{eqnarray}\label{HextP2}
  \mathcal{P} \mathbf{u}_i(z) &=& 
  \sum_{j}\left[\mathcal{P} \mathcal{H} \mathcal{P}\right]_{ij}^{-1}
  \left[-z\tau \mathcal{P} \mathbf{u}_j(z)+ \mathbf{f}_j/z\right]
  \nonumber \\ &&
  -\sum_{j,l} \left[\mathcal{P} \mathcal{H} \mathcal{P}\right]_{ij}^{-1}
  \mathcal{P} \mathcal{H}_{jl} \cdot
  \mathcal{Q}\mathbf{u}_l(z).
\end{eqnarray}
Then, we substitute the right-hand-side of Eq. \eqref{HextP2} into Eq. \eqref{HextQ1},
\begin{eqnarray}\label{HextQ2}
  && z\tau \mathcal{Q}\mathbf{u}_i(z)
  = - \sum_{j,l} \mathcal{Q} \mathcal{H}_{ij}\mathcal{P} 
  \left[\mathcal{P} \mathcal{H} \mathcal{P}\right]_{jl}^{-1}
  \left[-z\tau \mathcal{P} \mathbf{u}_l(z)+ \mathbf{f}_l/z\right]
  \nonumber \\ && 
  - \sum_j \left[\mathcal{Q} \mathcal{H}_{ij}
    - \sum_{l,m} \mathcal{Q} \mathcal{H}_{il}\mathcal{P} 
    \left[\mathcal{P} \mathcal{H} \mathcal{P}\right]_{lm}^{-1}
    \mathcal{P} \mathcal{H}_{mj}
    \right] \cdot \mathcal{Q}\mathbf{u}_j(z)
  \nonumber \\
\end{eqnarray}
Next, we solve Eq. \eqref{HextQ2} for $\mathcal{Q}\mathbf{u}_i$,
\begin{eqnarray}\label{HextQ3}
  && \mathcal{Q} \mathbf{u}_i(z) =
  \nonumber \\ &&
  -\sum_{l,m,j} \left\{ z\tau + \mathcal{Q} \mathcal{H}\mathcal{Q}
    -  \mathcal{Q} \mathcal{H}\mathcal{P}
    \left[\mathcal{P} \mathcal{H}\mathcal{P}\right]^{-1}
    \mathcal{P}\mathcal{H} \mathcal{Q} \right\}^{-1}_{il}
    \nonumber \\ && \times
    \mathcal{Q} \mathcal{H}_{lm}\mathcal{P}
    \left[\mathcal{P} \mathcal{H}\mathcal{P}\right]^{-1}_{mj} 
    \cdot\left[-z\tau \mathcal{P} \mathbf{u}_j(z)+ \mathbf{f}_j/z\right].
\end{eqnarray}
Finally, we substitute the right-hand-side of Eq. \eqref{HextQ3} into Eq. \eqref{HextP2}
and solve for $\mathcal{P}\mathbf{u}$, and in this way we obtain the following relation,
\begin{widetext}
\begin{eqnarray}\label{avedispl2}
  && \left\{
  z\tau + \left[\mathcal{P} \mathcal{H}\mathcal{P}\right]
  \left[\mathcal{P} \mathcal{H}\mathcal{P}
  + \mathcal{P} \mathcal{H}
  \mathcal{Q}
  \frac{1}{z\tau + \mathcal{Q} \mathcal{H} \mathcal{Q}
    -  \mathcal{Q} \mathcal{H}\mathcal{P}
    \left[\mathcal{P} \mathcal{H} \mathcal{P}\right]^{-1}
    \mathcal{P}\mathcal{H}  \mathcal{Q}}
  \mathcal{Q} \mathcal{H}\mathcal{P}\right]^{-1}
  \left[\mathcal{P} \mathcal{H}\mathcal{P}\right]\right\}\mathcal{P}\mathbf{u}
  = \mathbf{f}/z
\end{eqnarray}
\end{widetext}
where, once again, to simplify the notation we omitted summations over particles and
indices of the Hessian, and the displacement and force fields.
Equation \eqref{avedispl2} is the microscopic version of Eq. \eqref{affdispl}.
It is the alternative to Eq. \eqref{avedispl1} that we were looking for.

We note that structure of the matrix acting on $\mathcal{P}\mathbf{u}$ is very similar
to that obtained by Kawasaki \cite{Kawasaki}. This matrix was obtained by simple
manipulation of the same equations, Eqs. (\ref{HextP1}-\ref{HextQ1}), that were
used to obtain the standard relation between $\mathcal{P}\mathbf{u}$ and $\mathbf{f}/z$,
Eq. \eqref{avedispl1}.
In particular, although a person familiar with Kawasaki's analysis can clearly
see in Eq. \eqref{avedispl2} an object that could be called an ``irreducible Hessian'',
we arrived at Eq. \eqref{avedispl2} without introducing such a concept.

We emphasize that the above formal construction makes physical sense only if
operator $\mathcal{H} 
- \mathcal{H}\mathcal{P} \left[\mathcal{P} \mathcal{H} \mathcal{P}\right]^{-1}
\mathcal{P}\mathcal{H}$ is non-negative definite. We will return to this
question at the end of this section.

Again, to recover the alternative  expressions for the speeds of sound we need to
investigate the small wavevector limit of the second term at the left-hand-side
of Eq. \eqref{avedispl2} and then take the small $z\tau$ limit. The second term will
become the product of the squares of the speed of sound and the wavevector,
see Eq. \eqref{affdispl}.

We note that most objects involved in the small wavevector limit of the second term
at the left-hand-side of Eq. \eqref{avedispl2} were already discussed in the
context of the small wavevector limit of Eq. \eqref{avedispl1}. 
In Appendix \ref{appB} we discuss the new term,
$\mathcal{H}\mathcal{P}\left[\mathcal{P} \mathcal{H} \mathcal{P}\right]^{-1}
\mathcal{P}\mathcal{H}$, and additional steps needed to derive the small
limit of the second term at the left-hand-side of Eq. \eqref{avedispl2}.

Using results of Appendix \ref{appB} we obtain the following expressions
for the speeds of sound squared,
\begin{eqnarray}\label{c2long2}
  && c_{L}^2 =
  \frac{c_{LB}^4}
  {c_{LB}^2 + \frac{1}{N} \sum_{l,m} \hat{\mathbf{k}}\cdot\mathcal{W}_l(\hat{\mathbf{k}})
  \cdot\left[ \mathcal{H}-\delta_L\mathcal{H}\right]^{-1}_{lm}\cdot
  \mathcal{W}_m(\hat{\mathbf{k}})\cdot\hat{\mathbf{k}}},
  \nonumber \\
\end{eqnarray}
\begin{eqnarray}\label{c2trans2}
  && c_{T}^2 =\frac{c_{TB}^4}
  {c_{TB}^2 + \frac{1}{N}\sum_{l,m} \hat{\mathbf{e}}_T\cdot\mathcal{W}_l(\hat{\mathbf{k}})
  \cdot\left[ \mathcal{H}-\delta_T\mathcal{H}\right]^{-1}_{lm}\cdot
  \mathcal{W}_m(\hat{\mathbf{k}})\cdot\hat{\mathbf{e}}_T},
  \nonumber \\
\end{eqnarray}
where
\begin{eqnarray}\label{deltaLH}
  \delta_L\mathcal{H}_{lm} = \frac{1}{N}
  \mathcal{W}_l(\hat{\mathbf{k}})\cdot\hat{\mathbf{k}}
  c_{LB}^{-2} 
  \hat{\mathbf{k}}\cdot \mathcal{W}_m(\hat{\mathbf{k}}),
\end{eqnarray}
and
\begin{eqnarray}\label{deltaTH}
  \delta_T\mathcal{H}_{lm} = \frac{1}{N}
  \mathcal{W}_l(\hat{\mathbf{k}})\cdot\hat{\mathbf{e}}_T
  c_{TB}^{-2} 
  \hat{\mathbf{e}}_T\cdot \mathcal{W}_m(\hat{\mathbf{k}}).
\end{eqnarray}
Equations (\ref{c2long2}-\ref{c2trans2}) are our new
expressions for the speeds of sound squared, that we propose as
alternatives to standard expressions  (\ref{c2long1}-\ref{c2trans1}).

Finally, we need to examine the question of the non-negative definite
character of the matrices that enter into Eqs. (\ref{c2long2}-\ref{c2trans2}).
We note that, as discussed in Appendix \ref{appB},
matrices $\mathcal{H}-\delta_L\mathcal{H}$ and
$\mathcal{H}-\delta_T\mathcal{H}$ are obtained as $\mathbf{k}\to 0$ limits
of the following matrix
\begin{eqnarray}\label{posdef1}
  \mathcal{H}_{ij}(\mathbf{k})
  - \sum_{m,n} \mathcal{H}_{in}(\mathbf{k})\cdot\hat{\mathbf{e}}
  \left[\sum_{k,l}
    \hat{\mathbf{e}}\cdot\mathcal{H}_{kl}(\mathbf{k})\cdot\hat{\mathbf{e}}\right]^{-1}
  \hat{\mathbf{e}}\cdot\mathcal{H}_{mj}(\mathbf{k})
  \nonumber \\
\end{eqnarray}
with $\hat{\mathbf{e}}=\hat{\mathbf{e}}_L$ and $\hat{\mathbf{e}}=\hat{\mathbf{e}}_T$.
Next, we recall that the Hessian is non-negative definite. It follows
that the wavevector-dependent Hessian $\mathcal{H}(\mathbf{k})$
is also non-negative definite and thus can be written in the following way
in terms of its eigenvalues $\omega_a^2\ge 0$ and corresponding eigenvectors
$\boldsymbol{\mathcal{E}}_{ai}$, where $a$ labels eigenvectors.
\begin{eqnarray}\label{posdef2}
  \mathcal{H}_{ij}(\mathbf{k}) =
  \sum_a \omega_a^2 \boldsymbol{\mathcal{E}}_{ai} \boldsymbol{\mathcal{E}}_{aj}.
\end{eqnarray}
Using \eqref{posdef2} we can write a contraction of the matrix \eqref{posdef1}
with an arbitrary vector $\mathbf{a}_i$ as follows
\begin{eqnarray}\label{posdef3}
  &&
  \sum_a \left[\sum_i \omega_\alpha \boldsymbol{\mathcal{E}}_{ai} \cdot\mathbf{a}_i\right]^2
  \\ \nonumber && 
  - \frac{\left\{\sum_a \left[\sum_i \omega_a
    \boldsymbol{\mathcal{E}}_{ai} \cdot\mathbf{a}_i\right]
    \left[\sum_i \omega_a \boldsymbol{\mathcal{E}}_{ai} \cdot\hat{\mathbf{e}}\right]
  \right\}^2}
  {\sum_{b} \left[\sum_i \omega_b
      \boldsymbol{\mathcal{E}}_{bi} \cdot\hat{\mathbf{e}}\right]^2}.
\end{eqnarray}
Cauchy-Schwarz inequality implies that expression \eqref{posdef3} is non-negative
definite. Thus, matrices $\mathcal{H}-\delta_L\mathcal{H}$ and
$\mathcal{H}-\delta_T\mathcal{H}$ are also non-negative definite,
which makes our new expressions for speeds of sound squared,
Eqs. (\ref{c2long2}-\ref{c2trans2}) well defined. 

\section{Discussion}

We derived new exact formulae for elastic constants of zero-temperature
elastic solids. In contrast to standard expressions, our new formulae
are explicitly non-negative.

In practical numerical calculations, elastic constants of zero-temperature
elastic solids are determined by explicit deformations of these solids.
This procedure requires some care since one needs to balance between two opposite
goals: the need to impose small
enough deformation to ensure linear response and the need to impose large
enough deformation to generate statistically meaningful response signal.
However, it is simpler to implement than using standard exact formulae.
We expect that our new alternative formulae will also not be competitive with
the explicit deformation procedure. However, we hope that these formulae
will inspire future theoretical analyses of both elastic constants/speeds of sound
and of sound damping in zero-temperature elastic solids.

There were several theoretical analyses of sound propagation in zero-temperature
amorphous solids. In particular, the authors of Refs.
\cite{Ciliberti2003,Ganter2010,Grigera2011,Vogel2023} used different
diagrammatical analyses to analyze sound propagation, \textit{i.e.} speeds of
sound, and sound attenuation in zero-temperature amorphous solids. Our present
contribution suggests that resummations of classes of diagrams should be
attempted that would reproduce the structure of our new formulae for speeds
of sound. We hope to develop such theories in the near future. 

\section*{Acknowledgments}

I thank Florian Vogel and Matthias Fuchs for inspiring discussions
on possibly spurious instabilities predicted by approximate theories
of sound propagation in amorphous materials. 
I also thank them and Elijah Flenner for comments on the manuscript.
Most of the work described in this note was done when I was visiting
\'{Ecole} Normale Sup\'{e}rieure de Paris. I thank my colleagues there for their
hospitality.
The visit to ENS was partially 
supported by the Simons Foundation Grant 454955 (to Francesco Zamponi).
I gratefully acknowledge the support of NSF Grant No.~CHE 2154241. 

\appendix
\section{Analysis of the third term in Eq. \eqref{avedispl1}}\label{appA}

The third term at the left-hand-side of Eq. \eqref{avedispl1} involves
``vertices'' $\mathcal{P} \mathcal{H}_{il}\mathcal{Q}$ and
$\mathcal{Q}\mathcal{H}_{mj}\mathcal{P}$ and the inverse of the
projected Hessian $\left[z\tau + \mathcal{Q} \mathcal{H} \mathcal{Q} \right]_{lm}^{-1}$.
In the small wavevector limit the vertices are proportional to the
magnitude of the wavevector and can be expressed in terms of
tensorial field $\mathcal{W}_j$,
\begin{eqnarray}\label{W}
  \mathcal{W}_j(\hat{\mathbf{k}})
  = \sum_{l\neq j} \frac{\partial^2 V(R_{jl})}{\partial \mathbf{R}_j^2}
  \hat{\mathbf{k}}\cdot\mathbf{R}_{jl}.
\end{eqnarray}
The action of the vertices on an arbitrary vector $\mathbf{a}_i$ defined at each
inherent structure position reads,
\begin{eqnarray}\label{QHP1}
  && \sum_j \mathcal{Q}\mathcal{H}_{ij}\mathcal{P}\mathbf{a}_j
  = - i\left[ k \mathcal{W}_i(\hat{\mathbf{k}}) +o(k)\right] \cdot\mathcal{P}\mathbf{a}_i ,
\end{eqnarray}
\begin{eqnarray}\label{PHQ1}
  && \sum_j \mathcal{P}\mathcal{H}_{ij}\mathcal{Q}\cdot\mathbf{a}_j
  =i \left[ k \mathcal{P} \mathcal{W}_i(\hat{\mathbf{k}}) + o(k) \right]
  \cdot \mathbf{a}_i .
\end{eqnarray}

Tensorial field $\mathcal{W}_j$ is closely related to 
vector field $\boldsymbol{\Xi}_{j,\beta\delta}$ introduced by Lema\^{i}tre and C. Maloney
\cite{LemaitreJSP},
\begin{eqnarray}\label{Xi}
\boldsymbol{\Xi}_{j,\beta\delta} =
-\sum_{l\neq j} \frac{\partial^2 V(R_{jl})}{\partial R_{j\beta}\partial \mathbf{R}_{j}}
R_{jl\delta}.
\end{eqnarray}
Vector field $\boldsymbol{\Xi}_{j,\beta\delta}$
describes forces due to an affine deformation. Specifically,
$\boldsymbol{\Xi}_{j,\beta\delta}$ is proportional
to the force on particle $j$
resulting from a deformation along the $\beta$ direction that linearly depends
on the $\delta$ coordinates.
The $\alpha\beta$ component of $\mathcal{W}_j$ can be expressed
in terms of contraction of $\alpha$ component of vector
$\boldsymbol{\Xi}_{j,\beta\gamma}$ with $\hat{\mathbf{k}}$,
\begin{eqnarray}\label{WXi}
  \mathcal{W}_{j\alpha\beta}(\hat{\mathbf{k}})
  = \sum_{l\neq j} \frac{\partial^2 V(R_{jl})}{\partial R_{j\alpha}\partial R_{j\beta}}
R_{jl\delta} \hat{k}_\delta
\end{eqnarray}
where the Einstein summation convention over repeated Greek indices is adopted.

It follows that if we keep only the leading terms in the magnitude of the
wavevector in the vertices, we can re-write the third term
at the left-hand-side of Eq. \eqref{avedispl1} as
\begin{eqnarray}\label{QHQ1}
  && k^2 \mathcal{P} \mathcal{W}_i(\hat{\mathbf{k}})\cdot
  \left[z\tau + \mathcal{Q}\mathcal{H}\mathcal{Q}\right]^{-1}_{ij}\cdot
  \mathcal{W}_j(\hat{\mathbf{k}})\cdot\mathcal{P} \mathbf{u}_j
  \nonumber \\ &=&
  k^2 e^{-i\mathbf{k}\cdot\mathbf{R}_i} \hat{\mathbf{e}} \frac{1}{N}
  \sum_{lm} \hat{\mathbf{e}}\cdot
  \mathcal{W}_l(\hat{\mathbf{k}})\cdot
  \left[z\tau + \mathcal{Q}_1 \mathcal{H}(\mathbf{k})\mathcal{Q}_1\right]^{-1}_{lm}
  \nonumber \\ && \cdot
  \mathcal{W}_m(\hat{\mathbf{k}})\cdot\hat{\mathbf{e}}
  \frac{1}{N}\sum_j e^{i\mathbf{k}\cdot\mathbf{R}_j}  \hat{\mathbf{e}}\cdot\mathbf{u}_j
  \nonumber \\ &=&
  k^2 \frac{1}{N} \sum_{l,m} \hat{\mathbf{e}}\cdot
  \mathcal{W}_l(\hat{\mathbf{k}})\cdot
  \left[z\tau + \mathcal{Q}_1 \mathcal{H}(\mathbf{k})\mathcal{Q}_1\right]^{-1}_{lm}\cdot
  \mathcal{W}_m(\hat{\mathbf{k}})\cdot\hat{\mathbf{e}} \mathcal{P}\mathbf{u}_i
  \nonumber \\
\end{eqnarray}
where we introduced wavevector-dependent Hessian matrix $\mathcal{H}(\mathbf{k})$,
\begin{eqnarray}\label{Hk}
  \mathcal{H}_{il}(\mathbf{k}) =  \mathcal{H}_{il}
  e^{i\mathbf{k}\cdot\left(\mathbf{R}_i-\mathbf{R}_l\right)},
\end{eqnarray}
and a new (orthogonal) projection operator $\mathcal{Q}_1$ which acts on an arbitrary
vector $\mathbf{a}_i$ as follows
\begin{eqnarray}\label{Q1def}
  \mathcal{Q}_1\mathbf{a}_i = \mathbf{a}_i -
  \hat{\mathbf{e}} N^{-1}\sum_j \hat{\mathbf{e}} \cdot \mathbf{a}_j.
\end{eqnarray}

Finally, we show that in the small wavevector limit projection operators $\mathcal{Q}_1$
in $\sum_{l,m}  \hat{\mathbf{e}}\cdot
\mathcal{W}_l(\hat{\mathbf{k}})
\left[z\tau + \mathcal{Q}_1 \mathcal{H}(\mathbf{k})\mathcal{Q}_1\right]^{-1}_{lm}
\mathcal{W}_m(\hat{\mathbf{k}})\cdot\hat{\mathbf{e}}$ can be dropped. To this end we
use the procedure described by Ernst and Dorfman \cite{ErnstDorfman}. We use
the following operator identity
\begin{eqnarray}\label{ident1}
  \frac{1}{z\tau + \mathcal{H}(\mathbf{k})\mathcal{Q}_1} &=&
  \frac{1}{z\tau + \mathcal{H}(\mathbf{k})}
  \\ \nonumber &&
  + \frac{1}{z\tau + \mathcal{H}(\mathbf{k})\mathcal{Q}_1}
  \mathcal{H}(\mathbf{k})\mathcal{P}_1
  \frac{1}{z\tau + \mathcal{H}(\mathbf{k})},
  \end{eqnarray}
where projection operator $\mathcal{P}_1$ which acts on an arbitrary
vector $\mathbf{a}_i$ as follows
\begin{eqnarray}\label{P1def}
  \mathcal{P}_1\mathbf{a}_i =  \hat{\mathbf{e}} N^{-1}
  \sum_j \hat{\mathbf{e}} \cdot \mathbf{a}_j.
\end{eqnarray}
Using Eq. \eqref{ident1} and recalling that
$\mathcal{W}_l\mathcal{Q}_1= \mathcal{W}_l$ we re-write $\sum_{l,m}  \hat{\mathbf{e}}\cdot
\mathcal{W}_l(\hat{\mathbf{k}})\cdot
\left[z\tau + \mathcal{Q}_1 \mathcal{H}(\mathbf{k})\mathcal{Q}_1\right]^{-1}_{lm}\cdot
\mathcal{W}_m(\hat{\mathbf{k}})\cdot\hat{\mathbf{e}}$ as follows,
\begin{eqnarray}\label{ident2}
  && \sum_{l,m}\hat{\mathbf{e}}\cdot\mathcal{W}_l(\hat{\mathbf{k}})\cdot
  \left[z\tau + \mathcal{H}(\mathbf{k})\mathcal{Q}_1\right]^{-1}_{lm}\cdot
  \mathcal{W}_m(\hat{\mathbf{k}})\cdot\hat{\mathbf{e}}
  \\ \nonumber  &=& 
  \sum_{l,m}\hat{\mathbf{e}}\cdot\mathcal{W}_l(\hat{\mathbf{k}})\cdot
  \left[z\tau + \mathcal{H}(\mathbf{k})\right]^{-1}_{lm}\cdot
  \mathcal{W}_m(\hat{\mathbf{k}})\cdot\hat{\mathbf{e}}
  \\ \nonumber &&
  + \sum_{l,m,k,n}\hat{\mathbf{e}}\cdot\mathcal{W}_l(\hat{\mathbf{k}})\cdot
  \left[z\tau + \mathcal{H}(\mathbf{k})\mathcal{Q}_1\right]^{-1}_{lk}\cdot
  \mathcal{H}_{kn}(\mathbf{k})\cdot
  \\ \nonumber && \times
  \mathcal{P}_1
  \left[z\tau + \mathcal{H}(\mathbf{k})\right]^{-1}_{nm}\cdot
  \mathcal{W}_m(\hat{\mathbf{k}})\cdot\hat{\mathbf{e}}.
  \end{eqnarray}
In the second term at the right-hand-side of Eq. \eqref{ident2} we note that
\begin{eqnarray}\label{ident2a}
  \sum_n \mathcal{H}_{kn}(\mathbf{k})\mathcal{P}_1 =
  \left[-ik\mathcal{W}_k(\hat{\mathbf{k}}) + o(k)\right]\mathcal{P}_1.
\end{eqnarray}
Furthermore, we note that
\begin{eqnarray}\label{ident2b}
  &&
  z\tau N^{-1}\sum_{n,m}\hat{\mathbf{e}}\cdot
  \left[z\tau + \mathcal{H}(\mathbf{k})\right]^{-1}_{nm} \cdot
  \mathcal{W}_m(\hat{\mathbf{k}})\cdot\hat{\mathbf{e}} 
  \\ \nonumber &=&
  -i k N^{-1}\sum_{n,m}\hat{\mathbf{e}}\cdot\mathcal{W}_n(\hat{\mathbf{k}})\cdot
  \left[z\tau + \mathcal{H}(\mathbf{k})\right]^{-1}_{nm} \cdot
  \mathcal{W}_m(\hat{\mathbf{k}})\cdot\hat{\mathbf{e}} + O(k)
\end{eqnarray}
Combining Eqs. (\ref{ident2}-\ref{ident2b}) we obtain
\begin{eqnarray}\label{ident3}
  && \sum_{l,m}\hat{\mathbf{e}}\cdot\mathcal{W}_l(\hat{\mathbf{k}})\cdot
  \left[z\tau + \mathcal{H}(\mathbf{k})\right]^{-1}_{lm}\cdot
  \mathcal{W}_m(\hat{\mathbf{k}})\cdot\hat{\mathbf{e}}
  \\ \nonumber &=&
  \frac{z\tau \sum_{l,m}\hat{\mathbf{e}}\cdot\mathcal{W}_l(\hat{\mathbf{k}})\cdot
    \left[z\tau + \mathcal{H}(\mathbf{k})\mathcal{Q}_1\right]^{-1}_{lm}\cdot
    \mathcal{W}_m(\hat{\mathbf{k}})\cdot\hat{\mathbf{e}}}
       {z\tau - k^2 \sum_{l,m}\hat{\mathbf{e}}\cdot\mathcal{W}_l(\hat{\mathbf{k}})\cdot
         \left[z\tau +  \mathcal{H}(\mathbf{k})\mathcal{Q}_1\right]^{-1}_{lm}\cdot
         \mathcal{W}_m(\hat{\mathbf{k}})\cdot\hat{\mathbf{e}}}.
\end{eqnarray}
Equation \eqref{ident3} implies that in the small wavevector limit we can
drop projections $\mathcal{Q}_1$ in the last line of Eq. \eqref{QHQ1}. Note that
in the same limit we have $\mathcal{H}(\mathbf{k})\to \mathcal{H}$.

The final issue concerns the small $z\tau$ limit of the expression
$\sum_{l,m} \hat{\mathbf{e}}\cdot\mathcal{W}_l(\hat{\mathbf{k}})\cdot
\left[z\tau + \mathcal{H}\right]^{-1}_{lm}\cdot
\mathcal{W}_m(\hat{\mathbf{k}})\cdot\hat{\mathbf{e}}$. We recall that
the Hessian has zero eigenvalues. However, the small $z\tau$ limit
of $\sum_{l,m} \hat{\mathbf{e}}\cdot\mathcal{W}_l(\hat{\mathbf{k}})\cdot
\left[z\tau + \mathcal{H}\right]^{-1}_{lm}\cdot
\mathcal{W}_m(\hat{\mathbf{k}})\cdot\hat{\mathbf{e}}$ is well
defined for the following reason. First, we note that 
\begin{eqnarray}\label{zero1}
  && \sum_{l,m} \hat{\mathbf{e}}\cdot\mathcal{W}_l(\hat{\mathbf{k}})\cdot
  \left[z\tau + \mathcal{H}\right]^{-1}_{lm}\cdot
  \mathcal{W}_m(\hat{\mathbf{k}})\cdot\hat{\mathbf{e}}
  \nonumber \\ &=&
  \sum_{l} \hat{\mathbf{e}}\cdot\mathcal{W}_l(\hat{\mathbf{k}})\cdot
  \mathcal{U}_l(\hat{\mathbf{k}})\cdot\hat{\mathbf{e}},
  \nonumber \\
\end{eqnarray}
where $\mathcal{U}_l(\hat{\mathbf{k}})\cdot\hat{\mathbf{e}}$ satisfies the
following equation
\begin{eqnarray}\label{zero2}
  \mathcal{W}_l(\hat{\mathbf{k}})\cdot\hat{\mathbf{e}}
  =
  \sum_m \left[z\tau+\mathcal{H}\right]_{lm}\cdot
  \mathcal{U}_m(\hat{\mathbf{k}})\cdot\hat{\mathbf{e}}.
\end{eqnarray}
We note that $\mathcal{W}_m(\hat{\mathbf{k}})\cdot\hat{\mathbf{e}}$
is orthogonal to the space spanned by the eigenvectors of
$\mathcal{H}$ and thus in the $z\tau\to 0$ limit the solution of
Eq. \eqref{zero2} is well defined. 

We conclude that in the small wavevector and small $z\tau$ limit the
last line of Eq. \eqref{QHQ1} can be written as
\begin{eqnarray}\label{QHQ2}
  k^2 \frac{1}{N} 
  \sum_{l,m} \hat{\mathbf{e}}\cdot\mathcal{W}_l(\hat{\mathbf{k}})\cdot
  \left[ \mathcal{H}\right]^{-1}_{lm}\cdot
  \mathcal{W}_m(\hat{\mathbf{k}})\cdot\hat{\mathbf{e}}.
\end{eqnarray}

Finally, we note that for an isotropic system tensorial quantity
\begin{eqnarray}\label{QHQ3}
  \sum_{l,m} 
  \mathcal{W}_l(\hat{\mathbf{k}})\cdot
  \left[ \mathcal{H}\right]^{-1}_{lm}\cdot
  \mathcal{W}_m(\hat{\mathbf{k}})
\end{eqnarray}
consists of two independent components, longitudinal and transverse.
These components are the contributions to the squares of the speeds of sound
due to non-affine effects. They are the differences between the actual speeds of sound
squared and their Born avalues,
\begin{eqnarray}\label{QHQ4}
  \Delta c_{L}^2 =
  - \hat{\mathbf{k}}\cdot \frac{1}{N} \sum_{l,m} 
  \mathcal{W}_l(\hat{\mathbf{k}})\cdot
  \left[ \mathcal{H}\right]^{-1}_{lm}\cdot
  \mathcal{W}_m(\hat{\mathbf{k}})\cdot \hat{\mathbf{k}},
  \nonumber \\
\end{eqnarray}
\begin{eqnarray}\label{QHQ5}
  \Delta c_{T}^2 =
  - \frac{1}{2}
  \left[\boldsymbol{1} - \hat{\mathbf{k}}\hat{\mathbf{k}}\right] : \frac{1}{N} \sum_{lm} 
  \mathcal{W}_l(\hat{\mathbf{k}})\cdot
  \left[ \mathcal{H}\right]^{-1}_{l,m}\cdot
  \mathcal{W}_m(\hat{\mathbf{k}}).
  \nonumber \\
\end{eqnarray}

\section{Small wavevector limit of the second term in Eq. \eqref{avedispl2}}
\label{appB}

We start with the analysis of term
$\mathcal{P} \mathcal{H}\mathcal{Q}\left\{z\tau + \mathcal{Q} \mathcal{H} \mathcal{Q}
-  \mathcal{Q} \mathcal{H}\mathcal{P}
\left[\mathcal{P} \mathcal{H} \mathcal{P}\right]^{-1}
\mathcal{P}\mathcal{H}  \mathcal{Q}\right\}^{-1}\mathcal{Q} \mathcal{H}\mathcal{P}$.
Following the transformations similar to those used in writing Eq. \eqref{QHQ1},
if we keep only the leading terms in the magnitude of the wavevector in 
``outside'' vertices $\mathcal{P} \mathcal{H}\mathcal{Q}$ and
$\mathcal{Q} \mathcal{H}\mathcal{P}$, we can re-write this term as
\begin{eqnarray}\label{QHQ2nd1}
  && k^2 \frac{1}{N} \sum_{l,m} \hat{\mathbf{e}}\cdot
  \mathcal{W}_l(\hat{\mathbf{k}})\cdot
  \left[z\tau + \mathcal{Q}_1 \mathcal{H}(\mathbf{k})\mathcal{Q}_1
    \right. \\ \nonumber && \left.
    -  \mathcal{Q}_1 \mathcal{H}(\mathbf{k})\mathcal{P}_1
    \left[\mathcal{P}_1 \mathcal{H}(\mathbf{k}) \mathcal{P}_1\right]^{-1}
    \mathcal{P}_1\mathcal{H}(\mathbf{k})  \mathcal{Q}_1\right]^{-1}_{lm}\cdot
  \mathcal{W}_m(\hat{\mathbf{k}})\cdot\hat{\mathbf{e}} \mathcal{P},
\end{eqnarray}
where wavevector-dependent Hessian is defined in Eq. \eqref{Hk} and projection
operators $\mathcal{P}_1$ and $\mathcal{Q}_1$ are defined in
Eqs. \eqref{P1def} and \eqref{Q1def}, respectively.

Next, we use an identity analogous to Eq. \eqref{ident1} to write
the following relation,
\begin{widetext}
\begin{eqnarray}\label{ident4}
  && \sum_{l,m} \hat{\mathbf{e}}\cdot
  \mathcal{W}_l(\hat{\mathbf{k}})\cdot
  \left[z\tau + \mathcal{H}(\mathbf{k})\mathcal{Q}_1
    -  \mathcal{H}(\mathbf{k})\mathcal{P}_1
    \left[\mathcal{P}_1 \mathcal{H}(\mathbf{k}) \mathcal{P}_1\right]^{-1}
    \mathcal{P}_1\mathcal{H}(\mathbf{k})  \mathcal{Q}_1\right]^{-1}_{lm}\cdot
  \mathcal{W}_m(\hat{\mathbf{k}})\cdot\hat{\mathbf{e}}
  \nonumber \\ &=&
  \sum_{l,m} \hat{\mathbf{e}}\cdot
  \mathcal{W}_l(\hat{\mathbf{k}})\cdot
  \left[z\tau + \mathcal{H}(\mathbf{k})
    -  \mathcal{H}(\mathbf{k})\mathcal{P}_1
    \left[\mathcal{P}_1 \mathcal{H}(\mathbf{k}) \mathcal{P}_1\right]^{-1}
    \mathcal{P}_1\mathcal{H}(\mathbf{k})\right]^{-1}_{lm}\cdot
  \mathcal{W}_m(\hat{\mathbf{k}})\cdot\hat{\mathbf{e}}
  \nonumber \\ && +
  \sum_{l,m,k,n} \hat{\mathbf{e}}\cdot
  \mathcal{W}_l(\hat{\mathbf{k}})\cdot
  \left[z\tau + \mathcal{H}(\mathbf{k})\mathcal{Q}_1
    -  \mathcal{H}(\mathbf{k})\mathcal{P}_1
    \left[\mathcal{P}_1 \mathcal{H}(\mathbf{k}) \mathcal{P}_1\right]^{-1}
    \mathcal{P}_1\mathcal{H}(\mathbf{k})\mathcal{Q}_1\right]^{-1}_{lk}\cdot
  \left[\mathcal{H}(\mathbf{k})-\mathcal{H}(\mathbf{k})\mathcal{P}_1
    \left[\mathcal{P}_1 \mathcal{H}(\mathbf{k}) \mathcal{P}_1\right]^{-1}
    \mathcal{P}_1\mathcal{H}(\mathbf{k})\right]_{kn}
  \nonumber \\ && \cdot
  \mathcal{P}_1\left[z\tau + \mathcal{H}(\mathbf{k})
    -  \mathcal{H}(\mathbf{k})\mathcal{P}_1
    \left[\mathcal{P}_1 \mathcal{H}(\mathbf{k}) \mathcal{P}_1\right]^{-1}
    \mathcal{P}_1\mathcal{H}(\mathbf{k})\right]^{-1}_{nm}\cdot
  \mathcal{W}_m(\hat{\mathbf{k}})\cdot\hat{\mathbf{e}}
\end{eqnarray}
\end{widetext}
Then we note that
\begin{eqnarray}\label{ident4a}
  \sum_n\left[\mathcal{H}(\mathbf{k})-\mathcal{H}(\mathbf{k})\mathcal{P}_1
    \left[\mathcal{P}_1 \mathcal{H}(\mathbf{k}) \mathcal{P}_1\right]^{-1}
    \mathcal{P}_1\mathcal{H}(\mathbf{k})\right]_{kn}\mathcal{P}_1 = 0,
  \nonumber \\
\end{eqnarray}
which implies that the second term at the right-hand-side of Eq. \eqref{ident4}
vanishes. It follows that projections $\mathcal{Q}_1$ in Eq. \eqref{QHQ2nd1} do not
contribute and can be dropped.

To complete the analysis of term
$\mathcal{P} \mathcal{H}\mathcal{Q}\left\{z\tau + \mathcal{Q} \mathcal{H} \mathcal{Q}
-  \mathcal{Q} \mathcal{H}\mathcal{P}
\left[\mathcal{P} \mathcal{H} \mathcal{P}\right]^{-1}
\mathcal{P}\mathcal{H}  \mathcal{Q}\right\}^{-1}\mathcal{Q} \mathcal{H}\mathcal{P}$
we need to consider the small wavevector limit of
$ \mathcal{H}(\mathbf{k})\mathcal{P}_1
\left[\mathcal{P}_1 \mathcal{H}(\mathbf{k}) \mathcal{P}_1\right]^{-1}
\mathcal{P}_1\mathcal{H}(\mathbf{k})$. We have
\begin{eqnarray}\label{HPPHPPH}
  && \sum_{i,j} \mathcal{H}_{ki}(\mathbf{k})\mathcal{P}_1
  \left[\mathcal{P}_1 \mathcal{H}(\mathbf{k}) \mathcal{P}_1\right]^{-1}_{jl}
  \mathcal{P}_1\mathcal{H}_{jn}(\mathbf{k})
  \nonumber \\
  &=& 
  \sum_{i} \mathcal{H}_{ki}\cdot\hat{\mathbf{e}} e^{i\mathbf{k}\cdot\mathbf{R}_{ki}}
  \nonumber \\ && \times
  \left[c_{LB}^2 k^2 \left(\hat{\mathbf{k}}\cdot\hat{\mathbf{e}}\right)^2
    +c_{TB}^2 k^2 \left(1-\left(\hat{\mathbf{k}}\cdot\hat{\mathbf{e}}\right)^2\right)
    \right]^{-1}
  \nonumber \\ && \times
  \frac{1}{N}
  \sum_j e^{i\mathbf{k}\cdot\mathbf{R}_{jn}}\hat{\mathbf{e}}\cdot\mathcal{H}_{jn}
  \nonumber \\ &=&
  \mathcal{W}_k(\hat{\mathbf{k}})\cdot\hat{\mathbf{e}}
  \left[c_{LB}^2 \left(\hat{\mathbf{k}}\cdot\hat{\mathbf{e}}\right)^2
      +c_{TB}^2 \left(1-\left(\hat{\mathbf{k}}\cdot\hat{\mathbf{e}}\right)^2\right)
      \right]^{-1}
  \nonumber \\ && \times
  \frac{1}{N}\hat{\mathbf{e}}\cdot \mathcal{W}_n(\hat{\mathbf{k}})
  + O(k).
\end{eqnarray}

Thus, the small wavevector limit of term
$\mathcal{P} \mathcal{H}\mathcal{Q}\left\{z\tau + \mathcal{Q} \mathcal{H} \mathcal{Q}
-  \mathcal{Q} \mathcal{H}\mathcal{P}
\left[\mathcal{P} \mathcal{H} \mathcal{P}\right]^{-1}
\mathcal{P}\mathcal{H}  \mathcal{Q}\right\}^{-1}\mathcal{Q} \mathcal{H}\mathcal{P}$
reads
\begin{eqnarray}\label{QHQ2nd2}
  && k^2 \frac{1}{N} \sum_{l,m} \hat{\mathbf{e}}\cdot
  \mathcal{W}_l(\hat{\mathbf{k}})\cdot
  \left[z\tau + \mathcal{H}
    -  \frac{1}{N}\mathcal{W}(\hat{\mathbf{k}})\cdot\hat{\mathbf{e}}
    \right. \\ \nonumber && \left. \times
    \left[c_{LB}^2 \left(\hat{\mathbf{k}}\cdot\hat{\mathbf{e}}\right)^2
      +c_{TB}^2 \left(1-\left(\hat{\mathbf{k}}\cdot\hat{\mathbf{e}}\right)^2\right)
      \right]^{-1}
    \right. \\ \nonumber && \left. \times
     \hat{\mathbf{e}}\cdot \mathcal{W}(\hat{\mathbf{k}})
     \right]^{-1}_{lm}\cdot
  \mathcal{W}_m(\hat{\mathbf{k}})\cdot\hat{\mathbf{e}} \mathcal{P}
\end{eqnarray}

To write down the expression for the second term at the left-hand-side
of Eq. \eqref{avedispl2} we need to include $\mathcal{P}\mathcal{H}\mathcal{P}$ terms.
The resulting formula is rather long. To simplify it a bit, 
we will write it for the case of the transverse deformation, \textit{i.e.} for
$\hat{\mathbf{e}}=\hat{\mathbf{e}}_T$,
\begin{widetext}
 \begin{eqnarray}\label{QHQ2nd3}
   && \left. \left[\mathcal{P} \mathcal{H}\mathcal{P}\right]
   \left\{\mathcal{P} \mathcal{H}\mathcal{P}
     + \mathcal{P} \mathcal{H}
     \mathcal{Q}
     \frac{1}{z\tau + \mathcal{Q} \mathcal{H} \mathcal{Q}
       -  \mathcal{Q} \mathcal{H}\mathcal{P}
       \left[\mathcal{P} \mathcal{H} \mathcal{P}\right]^{-1}
       \mathcal{P}\mathcal{H}  \mathcal{Q}}
     \mathcal{Q} \mathcal{H}\mathcal{P}\right\}^{-1}
   \left[\mathcal{P} \mathcal{H}\mathcal{P}\right]\right|_T
   \nonumber \\ &\to &
   k^2  c_{TB}^2 \left\{k^2 c_{TB}^2 + k^2 \frac{1}{N} \sum_{lm} \hat{\mathbf{e}}_T\cdot
   \mathcal{W}_l(\hat{\mathbf{k}})\cdot
   \left[z\tau + \mathcal{H}
     - \delta_T\mathcal{H}
     \right]^{-1}_{lm}\cdot
   \mathcal{W}_m(\hat{\mathbf{k}})\cdot\hat{\mathbf{e}}_T
   \right\}^{-1} c_{TB}^2k^2 \mathcal{P}
 \end{eqnarray}
\end{widetext}
 where $\delta_T\mathcal{H}$ reads
\begin{eqnarray}\label{deltaTH2}
  \delta_T\mathcal{H} = \frac{1}{N}
  \mathcal{W}(\hat{\mathbf{k}})\cdot\hat{\mathbf{e}}_T
  c_{TB}^{-2} 
  \hat{\mathbf{e}}_T\cdot \mathcal{W}(\hat{\mathbf{k}}).
\end{eqnarray} 
Finally, we substitute expression \eqref{QHQ2nd3} into Eq. \eqref{avedispl2},
take the small $z\tau$ limit and we obtain the new microscopic
version of Eq. \eqref{affdispl}, with 
the square of the transverse speed of sound given in
Sec. \ref{alter}, Eq. \eqref{c2trans2}. The equation for the longitudinal
speed of sound can be obtained by substituting
$\hat{\mathbf{e}}_T\to\hat{\mathbf{e}}_L\equiv\hat{\mathbf{k}}$.


\begin{thebibliography}{99} 

\bibitem{BornHuang} 
  M. Born and K. Huang, \textit{Dynamical Theory of Crystal Lattices}, (Clarendon,
  Oxford, 1966).
\bibitem{Lutsko1989}
  J.F. Lutsko, ``Generalized expressions for the calculation of elastic constants
  by computer simulation'', J. Apl. Phys. \textbf{65}, 2991 (1989).
\bibitem{LeonfortePRB}
  F. Leonforte, R. Boissi\`{e}re, A. Tanguy, J.P. Wittmer, and J.-L. Barrat,
  ``Continuum limit of amorphous elastic bodies. III. Three-dimensional systems'',
  Phys. Rev. B \textbf{72}, 224206 (2005).
\bibitem{LemaitreJSP} A. Lema\^{i}tre and C. Maloney,
  ``Sum Rules for the Quasi-Static and Visco-Elastic Response of Disordered Solids 
  at Zero Temperature'',
  J. Stat. Phys. \textbf{123}, 415 (2006).
\bibitem{Wittmer2002}
  J.P. Wittmer, A. Tanguy, J.-L. Barrat, and L.J. Lewis, ``Vibrations of
  amorphous, nanometric structures: When does continuum theory apply?''
  EPL \textbf{57}, 423 (2002).
\bibitem{Tanguy2002}
  A. Tanguy, J.P. Wittmer, F. Leonforte, and J.-L. Barrat,
  ``Continuum limit of amorphous elastic bodies: A finite-size study of
  low-frequency harmonic vibrations'', Phys. Rev. B \textbf{66}, 174205 (2002).\
\bibitem{KarmakarLP2010}
  S. Karmakar, E. Lerner and I. Procaccia,
  ``Athermal nonlinear elastic constants of amorphous solids'',
  Phys. Rev. E \textbf{82}, 026105 (2010).
\bibitem{HessKlein} W. Hess and R. Klein,
  ``Generalized hydrodynamics of systems of Brownian particles'',
  Adv. Physics \textbf{32} (1983) 173.
\bibitem{JaeckleEisinger} J. J\"ackle and S. Eisinger,
  ``A hierarchically constrained kinetic Ising model'',
  Z. Phys. B \textbf{84}, 115 (1991).
\bibitem{CHess} B. Cichocki and W. Hess,
  ``On the memory function for the dynamic structure factor
  of interacting Brownian particles'', Physica A \textbf{141}, 475 (1987).
\bibitem{Kawasaki}K. Kawasaki,
  ``Irreducible memory function for dissipative
  stochastic systems with detailed balance'', Physica A \textbf{215}, 61 (1995).
\bibitem{SzamelLoewen} G. Szamel and H. L\"{o}wen,
  ``Mode-coupling theory of the glass transition in colloidal systems'',
  Phys. Rev. A \textbf{44}, 8215 (1991).
\bibitem{SastryDeBStil}
  S. Sastry, P.G. Debenedetti and F.H. Stillinger,
  ``Signatures of distinct dynamical regimes in the energy landscape of a glass-forming
  liquid'', Nature \textbf{393}, 554 (1998).
\bibitem{OHern2003}
  C.S. O'Hern, L.E. Silbert, A.J. Liu and S.R. Nagel,
  ``Jamming at zero temperature and zero applied stress: The epitome of disorder'',
  Phys. Rev. E \textbf{68}, 011306 (2003).
\bibitem{ErnstDorfman}
  M.H. Ernst and J.R.Dorfman,
  ``Nonanalytic Dispersion Relations for Classical Fluids. II. The General Fluid'',
  J. Stat. Phys. \textbf{12}, 311 (1975).
\bibitem{Ciliberti2003} S. Ciliberti, T.S. Grigera, V. Martin-Mayor, G. Parisi, and
  P. Verrocchio, ``Brillouin and boson peaks in glasses from
  vector euclidean random matrix theory''
  J. Chem. Phys. \textbf{119}, 8577 (2003).
\bibitem{Ganter2010}  C. Ganter and W. Schirmacher,
  ``Rayleigh scattering, long-time tails, and the harmonic spectrum of topologically
  disordered systems'', Phys. Rev. B \textbf{82}, 094205 (2010).
\bibitem{Grigera2011} T.S. Grigera, V. Martin-Mayor, G. Parisi, P. Urbani, and
  P. Verrocchio, ''On the high-density expansion for euclidean random matrices'',
  J. Stat. Mech. \textbf{2011}, P02015 (2011).
\bibitem{Vogel2023} F. Vogel and M. Fuchs,
  ``Vibrational phenomena in glasses at low temperatures captured by field theory of
  disordered harmonic oscillators'', arXiv:2211.10891.
\end{thebibliography}
\end{document}